\documentclass[useAMS,usenatbib,usegraphicx]{mn2e}

\title[Radiation from hot bare strange stars]
{Radiation from hot bare strange stars}
\author[A.G. Aksenov et al.]
{A.G.~Aksenov,$^{1,2}$ M.~Milgrom,$^1$ and V.V.~Usov,$^1$ \\
$^1$Department of Condensed Matter Physics, 
Weizmann Institute,
Rehovot 76100, Israel\\
$^2$Institute of
Theoretical and Experimental Physics, 
B.~Cheremushkinskaya, 25,
Moscow 117259, Russia}

\date{Accepted xxxxxx. Received 2003}

\pagerange{\pageref{firstpage}--\pageref{lastpage}} \pubyear{2002}

\begin{document}

\maketitle

\label{firstpage}

\begin{abstract}
We present the results of numerical simulations of stationary,
spherically outflowing, $e^\pm$ pair winds, with total
luminosities of $L=10^{35}- 10^{42}$ ergs~s$^{-1}$. These results
have direct relevance to the emission from hot, bare, strange
stars, which are thought to be powerful sources of pairs created by
the Coulomb barrier at the quark surface. The spectra of emergent
photons and pairs are calculated. For $L>2\times 10^{35}$
ergs~s$^{-1}$, photons dominate the emerging emission. As $L$
increases from $\sim 10^{35}$ to $10^{42}$ ergs~s$^{-1}$, the mean
photon energy decreases from $\sim 400-500$ keV to 40 keV, while
the spectrum changes in shape from a wide annihilation line to
being
nearly blackbody with a high energy ($> 100$ keV) tail. Such a
correlation of the photon spectrum with the luminosity, together
with the fact that super-Eddington luminosities can be achieved,
might be a good observational signature of hot, bare, strange
stars.
\end{abstract}


\section{Introduction}
Strange stars made entirely of deconfined quarks have long been
proposed as a possible alternative to neutron stars (e.g., Witten
1984; Alcock, Farhi, \& Olinto 1986a; Haensel, Zdunik, \&
Schaeffer 
1986). Because strange quark matter (SQM) at the surface of a bare
strange star is bound by strong interactions, not by gravity, such
a star can radiate with luminosities greatly exceeding the
Eddington limit. Luminosities as high as $10^{52}$~ergs~s$^{-1}$
can be radiated briefly after the formation of the star, when the
surface temperature is $T_{\rm S}\sim 10^{11}$~K (Alcock et al.
1986a; Chmaj, Haensel, \& Slomi\'nski 1991; Usov 1998,
2001a). Some 10 seconds after formation, $T_{\rm S}$ drops to less
than $10^{10}$~K (Page \& Usov 2002). At such temperatures--which
may also prevail later in the life of the star due to episodes of
reheating--electron-positron ($e^\pm$) pairs created by the
Coulomb barrier at the surface take up most of the thermal
emission from the SQM surface (Usov 2001a). Some of these pairs
annihilate into photons, thus forming a pair-photon wind that
flows away from the star.

For the typical radius of a strange star ($10^6$~cm), a
temperature of $T_{\rm S}\simeq 10^9-10^{10}$~K
gives an energy injection rate in pairs $\dot {\rm E}
\simeq 10^{43}-10^{49}$ ergs~s$^{-1}$ (Usov 2001a).
For such powerful winds the pair density near the surface is very
high, and the outflowing pairs and photons are very nearly
in thermal equilibrium almost up to the wind photosphere (e.g.,
Paczy\'nski 1990). The outflow may then be described fairly well
by relativistic hydrodynamics (Paczy\'nski 1986, 1990; Goodman
1986; Grimsrud \& Wasserman 1998; Iwamoto \& Takahara 2002). The
emerging emission consists mostly of photons, so $L_\gamma\simeq
\dot {\rm E}$. The photon spectrum is roughly a black body with a
temperature of $\sim 10^{10}(\dot {\rm E}/10^{49}~{\rm
ergs~s}^{-1})^{1/4}$~K. The emerging luminosity in $e^\pm$ pairs
is very small, $L_e=\dot {\rm E}-L_\gamma \sim 10^{-6} L_\gamma
\ll L_\gamma$. All this applies
roughly down to $\dot {\rm E} 
\sim 10^{42}-10^{43}$ ergs~s$^{-1}$.

\par
In contradistinction, for $\dot {\rm E} < 10^{42}$ ergs~s$^{-1}$
($T_{\rm S}< 9\times 10^8$ K), which is the region we explore,
the thermalization time for the pairs and photons is longer
than the escape time, and pairs and photons are not in thermal
equilibrium. In this Letter, we describe (for the first time to
our knowledge) the results of numerical calculations of the
characteristics of the emerging emission in pairs and photons in
stationary winds with energy injection rates $\dot {\rm E}=10^{35}-
10^{42}$ ergs~s$^{-1}$. The details of the set-up, code, processes
included, and the results for the wind structure, will be
described elsewhere (Aksenov, Milgrom, \& Usov 2003).
Here we give a brief account.

\section{Formulation of the problem}
At the moment of formation of
a strange star, the temperature in the stellar interior is expected
to be a few times $10^{11}$ K (Haensel, Paczy\'nski, \&
Amsterdamski 1991; Cheng \& Dai 2001). The neutrino luminosity of such 
a hot strange star is  up to $\sim 10^{54}$ ergs~s$^{-1}$.  
The rate of
neutrino-induced ejection of normal matter, which may be
present at the SQM surface (Alcock et al. 1986a;
Glendenning \& Weber 
1992), is
very high (e.g., Woosley \& Baron 1992; Levinson \& Eichler 
1993). Most probably, in a few seconds after the star forms, the 
normal-matter
envelope is blown away completely (Usov 2001a). 
The SQM surface remains exposed so as long as the
surface temperature is higher than $\sim 3\times 10^7$ K (Usov
1997). From the thermal evolution of young bare strange stars 
it follows that this may be the case for at least a few
hundred 
years after formation, and perhaps a few orders longer,
depending on the phase of SQM, convection, etc. (Page \&
Usov 2002)

We consider an $e^\pm$ pair wind that flows away from a hot, bare,
unmagnetized, strange star with a radius of $R=10^6$ cm. Pairs are
injected from time $t=0$ at a constant rate into the wind, which
is assumed spherical. This results in a time-dependent wind which
eventually becomes stationary.

We solve numerically the coupled, relativistic Boltzmann equations
for the pairs and photons (e.g., Mezzacappa \& Bruenn 1993;
Aksenov et al. 2003)
\begin{eqnarray}
  \frac{1}{c}\frac{\partial E_i}{\partial t}
 +{\mu\over r^2}\frac{\partial}{\partial r}
 (r^2 \beta_i E_i)
 +\frac{1}{r}\frac{\partial
 }{\partial\mu}
\left[(1-\mu^2)\beta_i E_i\right]
 \nonumber \\
 =\sum_q\left[\eta_i^q-\chi_i^q E_i\right], \label{Boltzmann}
\end{eqnarray}
where $E_i(\epsilon,\mu,r,t)={2\pi  \epsilon^3\beta_i f_i/ c^3}$
are the $\{{\bf r},\mu\,\epsilon\}$-phase-space energy densities of
$e^\pm$ pairs ($i=e$) and
photons ($i=\gamma$) ($f_i$ being the distribution functions);
$\mu =\cos \theta$, $-1\leq \mu\leq
1$, where $\theta$ is the angle between the radius-vector ${\bf
r}$ and the particle momentum ${\bf p}$; $\beta_\gamma =1$,
$\beta_e=v_e/c=\sqrt{1-(m_e c^2/\epsilon_e)^2}$; $\epsilon$ is the
particle energy; $\eta_i^q$ is the emission coefficient for the
energy production by a particle of type $i$ via the physical
process labelled by $q$; and, $\chi_i^q$ is the corresponding
absorption coefficient. The summation runs over all considered
physical processes. Gravity is neglected [but will be included in
our more detailed calculations (Aksenov et al. 2003)].
This is a good approximation  for the energy injection rates we
explore, because $L\simeq 10^{35}$ ergs~s$^{-1}$
is the Eddington luminosity for the pair
plasma, and above it, radiation-pressure forces dominate over
gravity.

 At the internal boundary, $r=R$, the input pair number flux depends
on the temperature $T_{\rm S}$ at the stellar surface alone, and
is taken as (Usov 2001a)
\begin{eqnarray}
F_e=10^{39}\left(\frac{\displaystyle
T_\mathrm{S}}{\displaystyle 10^9\,\mbox{K}}\right)^3
             \exp\left[-11.9\left(\frac{\displaystyle T_\mathrm{S}}
             {\displaystyle 10^9\,\mbox{K}}\right)^{-1}\right]
             \nonumber \\
             \times
             \left[\frac{\displaystyle\zeta^3\ln(1+2\zeta^{-1})}
                    {\displaystyle3(1+0.074\zeta)^3}
+
                    \frac{\displaystyle\pi^5\zeta^4}
                    {\displaystyle6(13.9+\zeta)^4}
             \right]{\rm cm}^{-2}{\rm s}^{-1},
\end{eqnarray}
where $\zeta=20(T_{\rm S}/10^9\,{\rm K})^{-1}$. Their energy
spectrum is thermal with temperature $T_{\rm S}$, and their
angular distribution is isotropic for $0\leq \mu \leq 1$. The
energy injection rate in $e^\pm$ pairs is then
\begin{equation}
\dot {\rm E}= 4\pi R^2[m_ec^2 +(3/2)k_{\rm B}T_{\rm S}] F_e\,,
\end{equation}
where $k_{\rm B}$ is the Boltzmann constant.

The stellar surface is assumed to be a perfect mirror for both
$e^\pm$ pairs and photons. At the external boundary ($r=r_{\rm
ext}$), the pairs and photons escape freely from the studied
region.

\begin{figure}
\includegraphics*[width=3.4in]{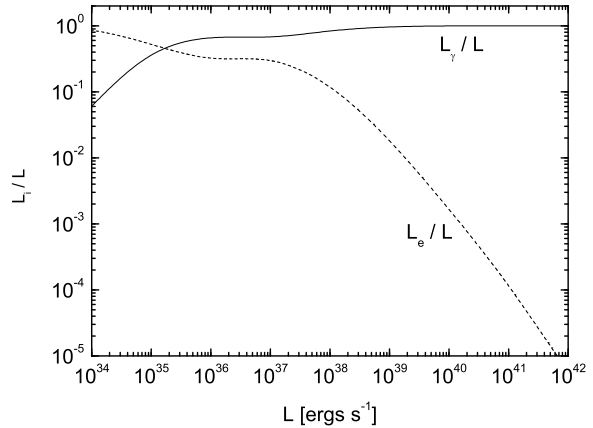}
 \caption{The partial luminosities in $e^\pm$
pairs ($L_e$) and photons ($L_\gamma$) as functions of the total
luminosity, $L$.}
 \label{fig1}
\end{figure}

\par
The thermal emission of pairs
used in our simulations [eq.~(2)]
is that expected in the
normal, unpaired, phase of SQM. 
However, it has been argued that SQM is a
color superconductor with a critical temperature of $\sim
10^{11}-10^{12}$ K (for a review, see Alford, Bowers, \& Rajagopal
2001; Rajagopal \& Wilczek 2001). It is expected that at density
much higher than the mean nuclear density, $\rho_{\rm n}\simeq
(2-3)\times 10^{14}$ g~cm$^{-3}$, the color superconductor is in
the ``Color-Flavor-Locked'' (CFL) phase, in which quarks of all
three flavors and three colors are paired in a single condensate.
If the density is only slightly higher than $\rho_{\rm n}$, SQM
may be in the ``2 color-flavor Superconductor'' (2SC) phase in
which only up and down quarks of two colors are paired while the
ones of the third color and the strange quarks of all three colors
are unpaired. In the CFL phase, SQM is electrically neutral and no
electrons are present. So, if at the surface the SQM is in the CFL
phase, there is no supercritical electric field and no $e^\pm$
pair emission. In the 2SC phase, electrons are present, and the
thermal emission of $e^\pm$ pairs from the surface is practically
the same as for normal SQM. Therefore, our simulations pertain
only to the last two cases.
\par
Although the injected pair plasma contains no radiation, as the
plasma moves outwards, photons are produced by pair annihilation.
Other two body processes that we include in our simulations are
M{\o}ller ($e^+e^+ \rightarrow e^+e^+ $ and $e^-e^-\rightarrow
e^-e^-$) and Bhaba ($e^+e^-\rightarrow e^+e^-$) scattering,
Comptom scattering ($\gamma e\rightarrow \gamma e$), and
photon-photon pair production ($\gamma\gamma\rightarrow e^+e^-$).
Two body processes do not change the total number of particles in
a system and thus cannot, in themselves, lead to thermal
equilibrium. For this reason, we include bremsstrahlung ($e
e\rightarrow e e\gamma$), double Compton scattering ($\gamma e
\rightarrow \gamma e \gamma$), and three quantum annihilation
($e^+e^-\rightarrow \gamma\gamma\gamma$), which change the
particle number, even though their cross-sections are
$\sim\alpha^{-1}\sim 10^2$ times smaller than those of the two
body processes ($\alpha$ is the fine-structure constant).

\section{Numerical results}
We give here the results for the properties of the emerging
radiation after stationarity is achieved, so the total wind
luminosity is equal to the energy injections rate:  $L=L_e
+L_\gamma=\dot {\rm E}$. We present results for different values 
of $L$,
which is the only free parameter. The corresponding surface
temperature $T_{\rm S}$ is found from equations (2) and (3).
Figure~\ref{fig1} shows the luminosity fractions in
pairs and photons. We see that for $L> L_*\simeq
2\times 10^{35}$ ergs~s$^{-1}$, the emerging emission
consists 
mostly of photons,
while $e^\pm$ pairs dominate for
$L < L_*$.
This simply reflects the fact that for $L < L_*$, the 
pair-annihilation time is longer than the escape time, and the injected
pairs remain mostly intact. At higher luminosities, most pairs
annihilate before escape, and reconversion into pairs is
inefficient, as  the mean energy of photons at the photosphere is
rather below the pair-creation threshold.

\begin{figure}
\includegraphics*[width=3.4in]{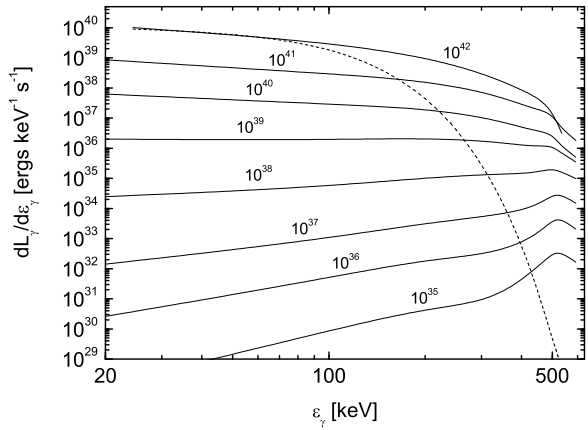}
 \caption{The energy spectra of emerging photons
for different values of the total luminosity, as marked on the
curves. Also shown, for $L=10^{42}$ ergs~s$^{-1}$, is
 the blackbody spectrum with the same energy
density as that of the photons at the photosphere of the
outflowing wind (dashed line).}
 \label{fig2}
\end{figure}

\begin{figure}
\includegraphics*[width=3.4in]{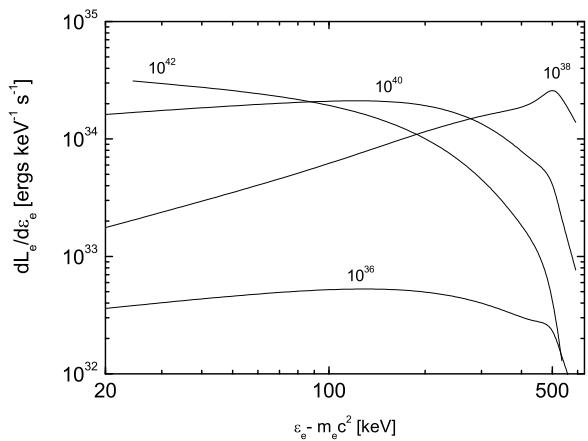}
 \caption{The energy spectra of emerging
electrons for different values of the total luminosity, as marked
on the curves.}
 \label{fig3}
\end{figure}

\begin{figure}
\includegraphics*[width=3.4in]{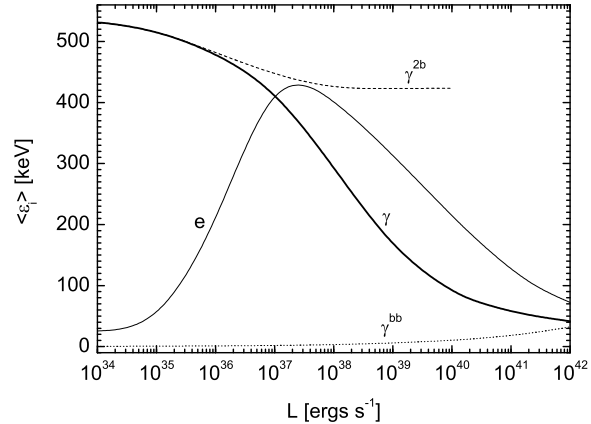}
 \caption{The mean energies of the emerging
photons (thick solid line) and electrons (thin solid line) as
functions of the total luminosity. For comparison we show as the
dotted line the mean energy of blackbody photons for the same
energy density as that of the photons at the photosphere of the
out-flowing wind. Also shown as the dashed line is
the mean energy
of the emerging photons in the case when only two body processes 
are
taken into account.}
 \label{fig4}
\end{figure}

Figures~\ref{fig2} and \ref{fig3} present the energy spectra of the 
emerging photons
and  pairs for different values of $L$. At low luminosities,
$L\sim 10^{35}-10^{37}$ ergs~s$^{-1}$, photons that form in
annihilation of $e^\pm$ pairs escape from the vicinity of the
strange star more or less freely, and the photon spectra resembles
a very wide annihilation line. The small decrease in mean photon
energy $\langle\epsilon_\gamma\rangle$ from $\sim 500$ keV at
$L\simeq 10^{35}$ ergs~s$^{-1}$ to $\sim 400$ keV at $L\simeq
10^{37}$ ergs~s$^{-1}$ occurs because of the energy transfer from
annihilation photons to $e^\pm$ pairs via Compton scattering (see
Fig.~\ref{fig4}). As a result of this transfer, the emerging $e^\pm$
pairs are heated up to the mean energy $\langle\epsilon_e\rangle
\simeq 400$ keV at $L\simeq 10^{37}$ ergs~s$^{-1}$. For $L>
10^{37}$ ergs~s$^{-1}$, changes in the particle number due to
three body processes are essential, and their role in
thermalization of the outflowing plasma increases with the
increase of $L$. We see in Figure 2 that, for $L= 10^{42}$
ergs~s$^{-1}$, the photon spectrum is near blackbody,
except for the presence of a high-energy tail at $\epsilon_\gamma
>100$ keV. At this luminosity, the mean energy
of the emerging photons is $\sim 40$ keV, while the mean energy of
the blackbody photons is $\sim 30$ keV (see Fig.~4). (The energy
resolution in our simulations is $\sim 20$ keV.)

For $L\simeq 10^{42}$ ergs~s$^{-1}$, the
emerging pair energy spectrum is close to a Maxwellian , while for
$L < 10^{40}$ ergs~s$^{-1}$ it deviates significantly from
it (see Fig.~3)
\section{Discussion}

We have identified certain characteristics of the expected
radiation from  hot, bare, strange stars that we hope will help
identify such stars, if they exist. The spectrum, we find, is
rather hard for the studied luminosity range. This makes such
stars amenable to detection and study by sensitive, high energy
instruments, such as INTEGRAL (e.g., Schoenfelder 2001), which is more
sensitive in this range than previous detectors.

As super-Eddington luminosities and, as we find, hard X-ray
spectra characterize the emission from bare, strange stars, soft
$\gamma$-ray repeaters (SGRs), which are the sources of short
bursts of hard X-rays with super-Eddington luminosities (up to
$\sim 10^{42}-10^{45}~{\rm ergs~s}^{-1}$), are reasonable
candidates for strange stars (e.g., Alcock, Farhi, \& Olinto
1986b; Cheng \& Dai 1998; Usov 2001b). The bursting activity of
SGRs may be explained by fast heating of the stellar surface up to
the temperature of $\sim (1-2)\times 10^9$~K and its subsequent
thermal emission (Usov 2001b,c). The heating mechanism may be
either 
impacts of comets onto bare strange stars (Zhang, Xu, \&
Qiao 2000; 
Usov 2001b) or fast decay of superstrong ($\sim 10^{14}-10^{15}$~G)
magnetic fields (Usov 1984; Thompson \& Duncan 1995; Heyl \& Kulkarni 
1998).
For typical luminosities of SGRs ($L\sim 10^{41}- 10^{42}$
ergs~s$^{-1}$),  the mean photons energy we find is $\sim 40$ keV
(see Fig.~4), which is consistent with observations of SGRs
(Hurley 2000).

Another important idiosyncrasy that we find is a strong
anti-correlation between spectral hardness and luminosity. While
at very high luminosities ($L>10^{42}-10^{43}$ ergs~s$^{-1}$) the spectral
temperature increases with luminosity as in blackbody radiation,
in the range of luminosities we studied, where thermal equilibrium
is not achieved, the expected correlation is opposite (see Fig. 4).
Such anti-correlations were, indeed, observed for SGR 1806-20 and
SGR 1900+14 where the burst statistic is high enough (e.g., Feroci
et al. 2001; Gogus et al. 2001; Ibrahim et al. 2001). This is
encouraging, but a direct comparison of this data with results
such as ours will require a more detailed analysis. In particular,
the effects of strong magnetic fields have to be included. The observed
periodic variations may be due to a rotation of a star with non-uniform
surface temperature, while our result apply for isotropic
emission. We hope to deal with this elsewhere.

\section*{Acknowledgments}
This work was supported by the Israel Science Foundation of
the Israel Academy of Sciences and Humanities.

\bsp

\label{lastpage}

\end{document}